# Frustrated Magnetism of the Maple-Leaf-Lattice Antiferromagnet MgMn$_3$O$_7$·3H$_2$O

Yuya Haraguchi*, Akira Matsuo, Koichi Kindo, Zenji Hiroi

The Institute for Solid State Physics, the University of Tokyo, Kashiwa, Chiba 277-8581, Japan

We present a novel hydrated layered manganate MgMn$_3$O$_7$·3H$_2$O as a maple-leaf-lattice (MLL) antiferromagnet candidate. The MLL is obtained by regularly depleting 1/7 of the lattice points from a triangular lattice so that the magnetic connectivity $z = 5$ and is thus intermediately frustrated between the triangular ($z = 6$) and kagomé ($z = 4$) lattices. In MgMn$_3$O$_7$·3H$_2$O, the Mn$^{4+}$ ions, carrying Heisenberg spin 3/2, form a regular MLL lattice in the quasi-two-dimensional structure. Magnetization and heat capacity measurements using a hydrothermally-prepared powder sample reveal successive antiferromagnetic transitions at 5 and 15 K. A high-field magnetization curve up to 60 T at 1.3 K exhibits a multi-step plateau-like anomaly. We discuss the unique frustration of the MLL antiferromagnet in which the chiral degree of freedom may play an important role.

## 1. Introduction

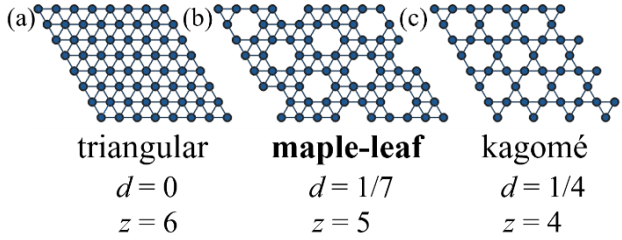

**Fig. 1** Comparison between three geometrically frustrated lattices in two dimensions: (a) triangular, (b) maple-leaf, and (c) kagomé lattices. Here, $d$ is the amount of depletion from the triangular lattice and $z$ is the magnetic connectivity for each lattice.

Geometrically frustrated magnets provide us with an opportunity to find exotic states such as the spin liquid state. In two dimensions, two types of magnetic lattice have been extensively studied: the triangular and kagomé lattices shown in Figs. 1(a) and (c). In a triangular lattice, each lattice point has $z = 6$ neighbors, while in a kagomé lattice, z is reduced to 4 as a result of the 1/4 depletion. On the other hand, there is another lattice between them, the maple-leaf lattice (MLL), which is obtained by depleting 1/7 of the points from the triangular lattice to give $z = 5$. When magnetic ions are located at the lattice points, there is only one kind of nearest-neighbor interaction in the triangular and kagomé lattices, while there are three in the MLL., as shown in Fig. 2: $J_d$, $J_t$, and $J_h$ connecting 2 sites; 3 sites in a triangle; and 6 sites in the hexagon, respectively.

For Heisenberg antiferromagnets without anisotropy or with easy-plane anisotropy in these frustrated lattices, a magnetic structure with a total spin of zero on each triangle is stable; this is often realized by arranging the 3 spins on the triangle at 120º to each other. In such a 120º structure, the vector chirality $\kappa$ is defined for each triangle as $\kappa = 2/(3\sqrt{3})(S_2 \times S_1 + S_3 \times S_2 + S_1 \times S_3)$. Following the convention that spins in the cross products are treated as if they are rotating counterclockwise around the triangle, the vector chirality points up or down normal to the plane in every triangle. Let us call the up and down $\kappa$ "positive" and "negative," respectively, as is widely used [5]. In the triangular lattice case, the vector chirality should be opposite on neighboring triangles, resulting in the unique staggered arrangement of the vector chirality. In contrast, for the kagomé lattice, the sign of the vector chirality on one triangle is not fixed by its surroundings due to the low connectivity that causes macroscopic degeneracy in the ground state and tends to destroy simple magnetic order. Therefore, the kagomé lattice is more frustrated than the triangular lattice.

In the MLL case, the vector chirality may emerge in a similar 120º spin order. Expected for the specific case where $J_d = J_t = J_h$ is a magnetic structure with 120°-rotated spins on the $J_t$ trimer, which are arranged such that two spins on the $J_d$ dimer are at a 90° angle to each other [6,7]. In this structure, the vector chiralities on the $J_t$ trimer are in a staggered arrangement. On the other hand, for the general case, the spin and chirality arrangements in the MLL antiferromagnet have not been clarified.

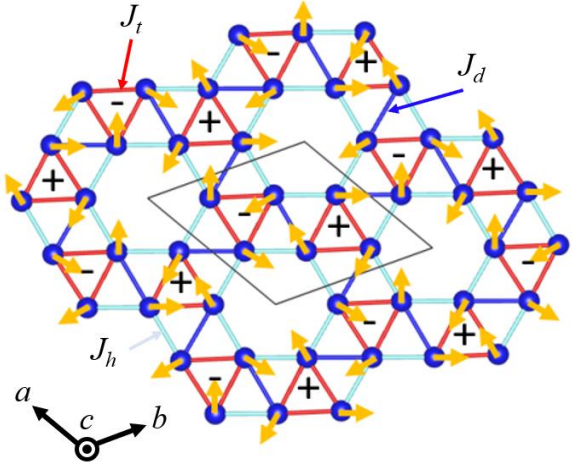

**Fig. 2** Magnetic network with the three kinds of magnetic interactions $J_d$, $J_t$, and $J_h$ in the maple-leaf lattice antiferromagnet. The yellow arrows represent a spin arrangement expected in the case of $J_d = J_t = J_h$ [5, 6], which may not be realized in actual magnets with nonequivalent interactions.

The MLL antiferromagnet may be an interesting frustrated lattice, but it has been studied less compared with the triangular and kagomé lattice antiferromagnets. The main reason for this is the lack of model compounds. So far, the compounds studied as MLL antiferromagnets have been the natural mineral Spangolite $Cu_6Al(SO_4)(OH)_{12}Cl_3·H_2O$ [8], and $[Mn_{3+x}O_7][Bi_4O_{4.5-y}]$ [9]. The former suffers from the presence of nonmagnetic impurities at the magnetic $Cu^{2+}$ sites, while the latter suffers from lattice distortion and extra magnetic $Mn^{4+}$ ions between the layers. Thus, a new model compound is required to uncover the properties of the MLL antiferromagnet.

In this paper, we report on the magnetic properties of the novel MLL antiferromagnet $MgMn_3O_7·3H_2O$. It is known as the natural mineral jianshuiite [11], which is isomorphic to chalcophanite $ZnMn_3O_7·3H_2O$ [12] and erneickelite $NiMn_3O_7·3H_2O$ [13]. However, the details of its crystal structure and magnetism have not yet been studied. We prepared a powder sample of $MgMn_3O_7·3H_2O$ following the hydrothermal method, and measured the magnetic susceptibility, heat capacity, and high-field magnetization. The compound exhibits successive phase transitions upon cooling and with an increasing magnetic field. We propose a spin model and possible magnetic structures from the viewpoint of the crystal structure. We also discuss the frustrated magnetism of the MLL antiferromagnet.

## 2. Experiments

A polycrystalline sample of $MgMn_3O_7·3H_2O$ was prepared using the hydrothermal method. A mixture of 0.15 g of $KMnO_4$ and 1.50 g of $Mg(NO_3)_2·6H_2O$ was put into a Pyrex beaker and placed in a stainless-steel vessel 30 ml in volume. The vessel was filled with 20 ml of $H_2O$, sealed, and heated at 250 °C for 24 h. The product was characterized by powder X-ray diffraction (XRD) in a diffractometer with Cu-$K\alpha$ radiation. The cell parameters and crystal structure were refined according to the Rietveld method using the RIETAN-FP v2.16 software [14]. The temperature dependence of the magnetization was measured for randomly-oriented powder samples under magnetic fields up to 7 T in a magnetic property measurement system (MPMS; Quantum Design). Since the $Mn^{4+}$ ion with the $d_3$ electron configuration has no orbital degree of freedom, an alignment of the powder in a magnetic field must not occur. The temperature dependence of the heat capacity was measured using the conventional relaxation method in a physical property measurement system (PPMS; Quantum Design). Magnetization curves up to approximately 60 T were measured using an induction method with a pulsed magnet at the International Mega Gauss Science Laboratory of the Institute for Solid State Physics at the University of Tokyo; because of the small anisotropy of the $Mn^{4+}$ spins and the magnetic field's short duration time of 5 ms, the random orientation of the power must be maintained during the experiment.

## 3. Results

### 3.1 Crystal Structure

The powder XRD pattern for the hydrothermally prepared sample is shown in Fig. 3(a).

All of the peaks are indexed to reflections based on the space group of $R\bar{3}$ with the lattice constants $a = 7.5293(4)$ Å and $c = 20.752(1)$ Å, which are similar to those of chalcophanite, $ZnMn_3O_7·3H_2O$ [12] and ernenickelite, $NiMn_3O_7·3H_2O$ [13]. The chemical composition was examined by energy dispersive X-ray spectrometry and it was found that Mn/Mg = 3.00(4). Thus, we had successfully synthesized $MgMn_3O_7·3H_2O$. In fact, our Rietveld refinement converges well with the expected structure. The refined crystallographic parameters are listed in Table 1, where $B$ is the isotropic thermal displacement parameter and WO denotes the oxygen in the $H_2O$ molecule. The values of $B$ at the Mg and WO sites are relatively large compared with those at the Mn and O sites, respectively. This may indicate the presence of disorder at these sites.

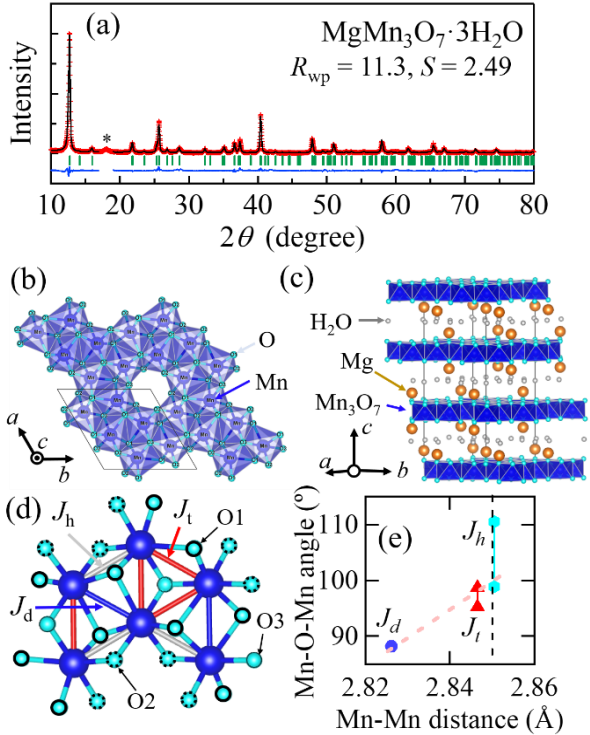

**Fig. 3** (a) XRD pattern of a powder sample of $MgMn_3O_7·3H_2O$. (b), (c) Crystal structures of $MgMn_3O_7·3H_2O$ viewed along the $c$ axis and the $ab$ plane, respectively. (d) Local atomic arrangements around the Mn ions with three nonequivalent magnetic interactions $J_d$, $J_t$, and $J_h$. (e) Mn–O–Mn angle versus the Mn–Mn distance for the three magnetic interactions. The two different Mn–O–Mn angles for $J_t$ ($J_h$) correspond to the two superexchange pathways via the Mn–O1–Mn and Mn–O3–Mn (Mn–O1–Mn and Mn–O2–Mn) bonds.

The crystal structure of $MgMn_3O_7·3H_2O$ is shown in Figs. 3(b) and (c). There is only one crystallographic site for the Mn atom, which is octahedrally coordinated by 6 oxygen atoms. The $MnO_6$ octahedra are linked by common edges to form a two-dimensional layer. Since 1/7 of the Mn atoms are regularly depleted from a triangular lattice, a $Mn_{6/7}O_2 = Mn_3O_7$ layer containing a MLL lattice made of $Mn^{4+}$ ions is generated. As shown in Fig. 3(c), the $Mn_3O_7$ layers are well separated from each other by a nonmagnetic block layer consisting of $Mg^{2+}$ ions located above and below the Mn-vacant positions of the $Mn_3O_7$ layer and crystallization water molecules. The Mn–Mn distance between the layers is ~6.5 Å, which is significantly larger than ~2.8 Å in the layer. Therefore, good magnetic two-dimensionality is expected for $MgMn_3O_7·3H_2O$.

**Table 1** Crystallographic parameters for $MgMn_3O_7·3H_2O$ (space group: $R\bar{3}$) determined by the Rietveld refinement of powder XRD data. The lattice parameters are $a = 7.5293(4)$ Å and $c = 20.752(1)$ Å. $B$ is the isotropic thermal displacement parameter. WO denotes the oxygen in the $H_2O$ molecule. Refinements converged with the reliable parameter $R_{wp} = 11.3$ %, and $S = 2.49$.

| Atom | Site | x | y | z | B(Å²) |
|---|---|---|---|---|---|
| Mn | 18f | 0.2375(4) | 0.0460(3) | 0.3324(1) | 0.54 |
| Mg | 6c | 0 | 0 | 0.1026(3) | 1.8 |
| O1 | 18f | 0.2364(1) | 0.2221(11) | 0.3759(4) | 0.67 |
| O2 | 18f | 0.2548(13) | 0.0535(16) | 0.0485(3) | 0.80 |
| O3 | 6c | 0 | 0 | 0.2876(5) | 0.65 |
| WO | 18f | 0.1820(14) | 0.2589(11) | 0.1607(4) | 1.3 |

The $Mn^{4+}$ ions form an MLL without distortion. There are three kinds of exchange paths between Mn neighbors, which correspond to the $J_d$, $J_t$, and $J_h$ of MLL. The Mn–Mn distance becomes larger in this order, as shown in Fig. 3(e), which means that the direct exchange interaction must decrease accordingly. Moreover, there are superexchange interactions via oxygen ions, as shown in Fig. 3(d), the magnitude of which should depend on the Mn–O–Mn bond angle. The two equivalent Mn–O1–Mn superexchange paths contribute the $J_d$ coupling, while the $J_t$ ($J_h$) coupling is mediated by the two superexchange paths Mn-O1-Mn and Mn–O3–Mn (Mn–O1–Mn and Mn–O2–Mn). The bond angles are approximately 90° for $J_d$, and 100–110° for $J_t$ and $J_h$, as shown in Fig. 3(e). The total magnitude of each exchange interaction is the sum of the direct and superexchange couplings, which we discuss later.

## 3.2 Magnetic Property

Figure 4 shows the temperature dependence of the magnetic susceptibility $M/H$ and its inverse for MgMn$_3$O$_7$·3H$_2$O measured at $\mu_0 H = 1$ T. A Curie–Weiss fitting of the inverse susceptibility at 200–300 K yields an effective magnetic moment $\mu_{eff} = 3.95\mu_B$ and Weiss temperature $\theta_W = -60.1$ K. The effective moment is close to the spin-only value for $S = 3/2$ ($g = 2.04$), indicating an isotropic Heisenberg spin with few spin-orbit interactions, as expected for the $d^3$ electron configuration. The large negative value of $\theta_W$ indicates predominantly antiferromagnetic interactions between the Mn$^{4+}$ spins. The average magnetic interaction $J = (J_d + 2J_t + 2J_h)/5$ is calculated to be $-9.62$ K in the mean field approximation; $\theta_W = 2/3zJS(S + 1)$ with $z = 5$ and $S = 3/2$.

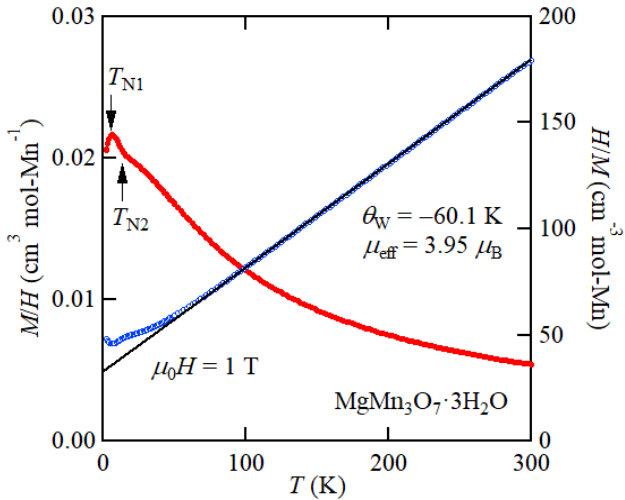

**Fig. 4** Temperature dependence of magnetic susceptibility $M/H$ and its inverse for MgMn$_3$O$_7$·3H$_2$O under a magnetic field of 1 T. The solid line on the inverse susceptibility indicates the result of Curie-Weiss fitting.

Below 50 K, the magnetic susceptibility becomes smaller than expected from the Curie–Weiss law; this indicated the development of an antiferromagnetic short-range order. At lower temperatures below 15 K, it starts to increase followed by a hump at 5 K. Fig. 5(a) expands the low-temperature region measured under magnetic fields from 0.01 to 7 T. At the lowest field of 0.01 T, $M/H$ exhibits two humps at $T_{N1} = 5$ K and $T_{N2} = 15$ K, as well as a thermal hysteresis between the zero-field-cooled and field-cooled data, suggesting successive magnetic transitions. These humps and hysteresis are suppressed by increasing the magnetic field and eventually merged into a broad 20 K peak at 7 T. Figure 5(b) shows the frequency dependence of the real part of the ac susceptibility $\chi'$. There are large and small peaks at approximately 10 and 20 K, respectively, both of which shift to higher temperatures and become smaller as the frequency increases. Thus, the observed magnetic anomalies may be associated with certain slow spin dynamics.

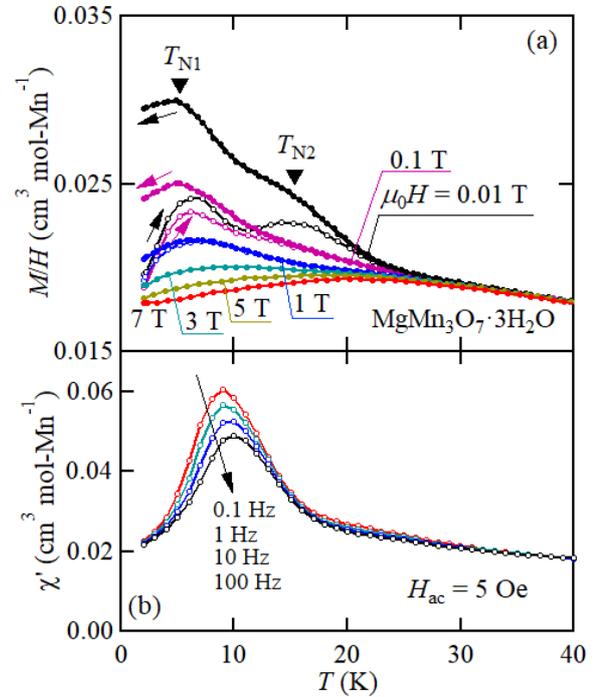

**Fig. 5** (a) Temperature dependences of $M/H$ measured for several magnetic field strengths. In each field, measurements were conducted upon heating after zero-field cooling, and then upon cooling, as shown by the arrows. (b) Temperature dependences of the real part of the ac susceptibility measured in oscillating magnetic fields of $H_{ac} = 5$ Oe at frequencies of 0.1, 1, 10, and 100 Hz.

Figure 6(a) shows the isothermal magnetization curves at 50, 4.2, and 1.8 K. The curve at 50 K is linear, while those below $T_{N2}$ show small spontaneous magnetizations with hystereses, indicating the presence of a weak ferromagnetic moment accompanied by the magnetic ordering. The inset of Fig. 6(b) shows the temperature evolution of the magnetic hysteresis loops $M_{down} - M_{up}$, where $M_{down}$ and

$M_{up}$ are the magnetizations measured in down- and up-sweeping fields, respectively. The magnetic hysteresis loop shrinks as the temperature increases and vanishes at 50 K. The main panel of Fig. 6(b) displays the temperature dependence of spontaneous magnetization $M_{sp}$, which is defined as the half value of $M_{down}-M_{up}$ at 0 T. Close to $T_{N2}$, $M_{sp}$ starts to increase and rapidly increases below $T_{N1}$, indicating that antiferromagnetic ordering with a weak ferromagnetic moment occurs at $T_{N1}$, and that another magnetic ordering with a larger ferromagnetic moment follows at $T_{N2}$. The slow spin dynamics observed in the ac magnetization measurement may not be due to a spin glass transition but related to the slow dynamics of weak ferromagnetic domains.

### 3.3. Thermodynamic property

Figure 7 shows the temperature dependence of the heat capacity divided by temperature $C/T$ of $MgMn_3O_7 \cdot 3H_2O$. There is a large hump at low temperatures, which consists of two broad peaks at approximately $T_{N1}$ and $T_{N2}$. Thus, the magnetic anomalies at $T_{N1}$ and $T_{N2}$ must be ascribed to bulk magnetic phase transitions. The broadened transitions may be intrinsic due to a certain frustration effects or due to an inhomogeneity of our sample. The inset of Fig. 7 shows the $C/T$ data measured under different magnetic fields. In contrast to the large magnetic field dependence of the magnetic susceptibility, the heat capacity is only slightly dependent on the magnetic field; this is consistent with the fact that the former is a result of the weak-ferromagnetic moment.

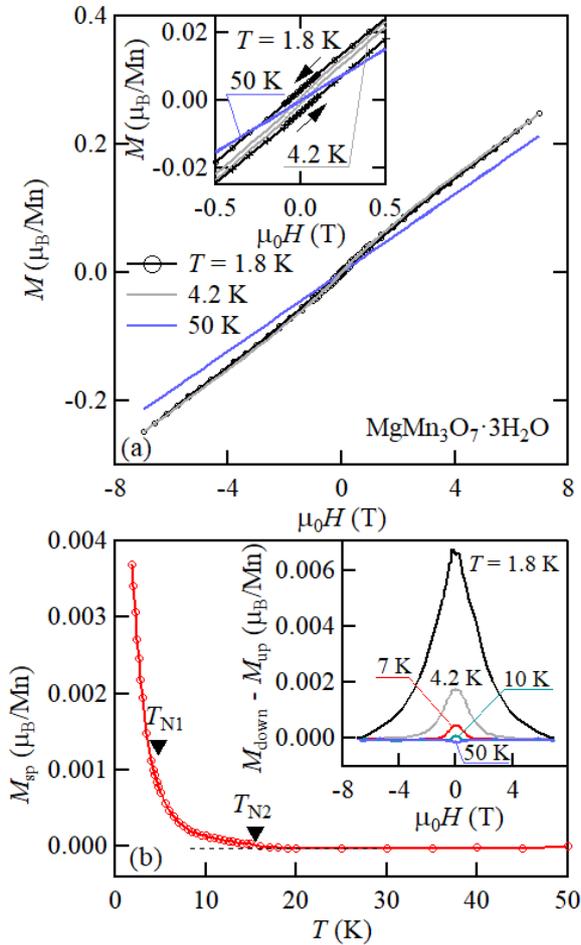

**Fig. 6** (a) Isothermal magnetization curves measured at 50, 4.2, and 1.8 K for $MgMn_3O_7 \cdot 3H_2O$. The inset shows an enlarged view approaching 0 T. (b) Temperature dependence of the spontaneous magnetization $M_{sp}$. The inset shows differences in the magnetization between the down- and up-sweep curves; $M_{down}-M_{up}$.

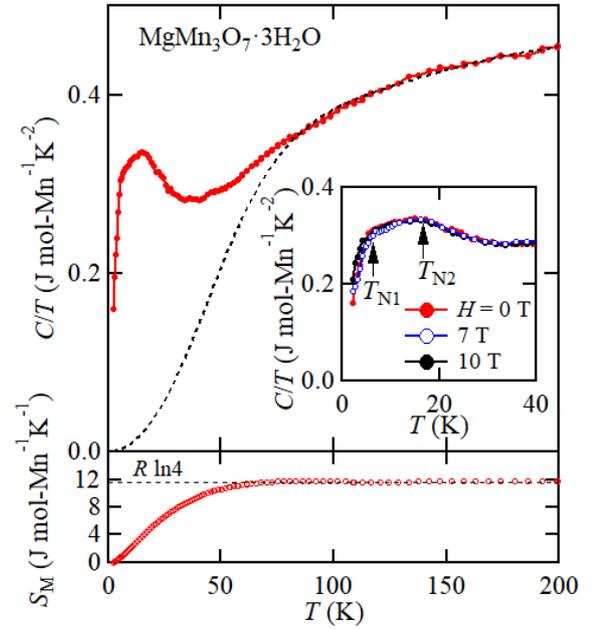

**Fig. 7** Temperature dependence of the heat capacity divided by temperature $C/T$ for $MgMn_3O_7 \cdot 3H_2O$. The dashed line represents lattice contributions estimated by fitting the data above 100 K, as described in the text. The inset shows $C/T$ data at $H = 0$, 7, and 10 T. The magnetic entropy $S_M$ is shown in the bottom panel.

To extract the magnetic contribution to the heat capacity, the lattice contribution was estimated by fitting the high-temperature data. Provided that it is the sum of Debye- and Einstein-type heat capacities, $C_D$ and $C_E$ respectively, the $C/T$ data above 100 K, where the magnetic heat capacity may be negligible, are fitted to the equation $C/T = 3R\{aC_D/T + (20/3 - a)C_E/T\}$, where $R$ is the gas constant and $a$ is the weight parameter. The best fit is shown by the dashed line with $a = 0.977(15)$, Debye temperature $\theta_D = 396(4)$ K, and Einstein temperature $\theta_E = 925(8)$ K. The magnetic contribution was obtained by subtracting this lattice contribution from the experimental data, and the magnetic entropy $S_M$ was calculated by integrating the magnetic $C/T$ with respect to $T$. The asymptotic value of $S_M$ at high temperatures coincides with the expected total magnetic entropy for spin 3/2, $R\ln 4 = 11.52$ J·mol$^{-1}$·K$^{-1}$, demonstrating that subtraction of the lattice contribution was valid. The $S_M$ reaches approximately 1.43 and 5.37 J·mol$^{-1}$·K$^{-1}$ at $T_{N1}$ and $T_{N2}$, which are 12 and 47 % of the total magnetic entropy, respectively. This indicates that a large part of the magnetic entropy was released by the development of short-range magnetic correlations above $T_{N2}$.

### 3.4 High field magnetization

To get further information about the magnetic ordering and magnetic interactions in MgMn$_3$O$_7$·3H$_2$O, magnetization measurements up to 60 T are conducted using pulsed magnetic fields, as shown in Fig. 8. In the low magnetic field regions, the $M$–$H$ curve at 1.3 K increases linearly, as evidenced by the constant $dM/dH$. At higher fields, the $M$–$H$ curve increases over 3 steps at $H_{s1} = 33$ T, $H_{s2} = 40$ T, and $H_{s3} = 51$ T, where $dM/dH$ shows peaks, and tends to saturate above 60 T. The maximum magnetization is much smaller than the saturation moment $M_s = 3.06\mu_B$ for $S = 3/2$ and $g = 2.04$, and is close to $(3/4)M_s$. All of the observed anomalies become smaller at 4.2 K, indicating the presence of a series of magnetic-field induced phase transitions at low temperatures.

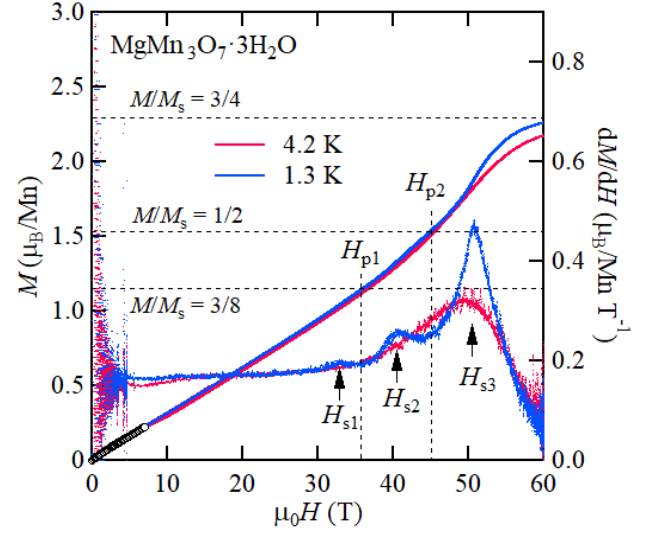

Fig. 8  Magnetization curves M–H and their derivatives measured in pulsed fields up to 60 T at 1.3 and 4.2 K for MgMn3O7·3H2O. The open circles below 7 T are measured in static fields, which are used to calibrate the high-field data. The vertical broken lines represent the local minimum point of dM/dH, indicating the presence of magnetization plateaus. The horizontal lines indicate the value of approximately 3/8, 1/2, and 3/4 magnetization saturation. The arrows indicate the critical fields.

## 4. Discussion

### 4.1 Magnetic interactions

MgMn$_3$O$_7$·3H$_2$O is a MLL spin-3/2 Heisenberg antiferromagnet. First, we consider the magnitude of the magnetic exchange interactions based on the crystal structure. There are three non-equivalent

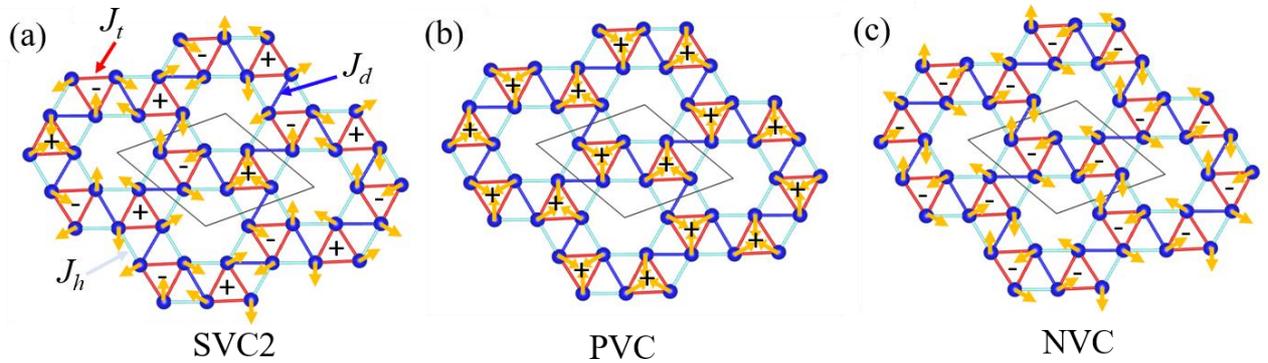

**Fig. 9** Possible magnetic structures for the MLL antiferromagnet in the case where $J_d > J_t > J_h$: (a) SVC2 (staggered vector chirality), (b) PVC (positive vector chirality), and (c) NVC (negative vector chirality) orders. The marks + and − on each triangle represent the up and down vector chirality, respectively.

Mn–Mn magnetic exchange interactions $J_d$, $J_t$, and $J_h$. As shown in Fig. 3, $J_d$ has a short Mn–Mn distance $d_{Mn–Mn}$ of 2.826 Å and a Mn–O–Mn bond angle of approximately 90º, while $J_t$ and $J_h$ have longer Mn–Mn distances of ~2.85 Å and large Mn-Mn bond angles of 95–100º.

In general, the direct exchange interaction between $d^3(t_{2g}^3)$ ions such as $Mn^{4+}$ is antiferromagnetic, and its value is inversely proportional to the bond length [14,15]. On the other hand, according to the Kanamori–Goodenough rule, the superexchange interaction between $d^3$ spins is ferromagnetic at bond angles close to 90º and becomes antiferromagnetic as the bond angle increases [16]. It is known, from previous ESR study of manganese spinel compounds, that the $Mn^{4+}$–$Mn^{4+}$ coupling in a pair of edge-sharing $MnO_6$ molecules changes from antiferromagnetic to ferromagnetic at $d_{Mn–Mn} = 2.85$ Å [17]. This is because direct exchange interactions between half-filled $t_{2g}$ orbitals are much stronger at short Mn–Mn distances and the super exchange interactions prevail as the distance decreases and the bond angle increases. For $MgMn_3O_7 \cdot 3H_2O$ with $d_{Mn–Mn} < 2.85$ Å all of the couplings must be antiferromagnetic and their magnitudes depend on the Mn–Mn distances. Thus, we conclude that $J_d$ is significantly larger than $J_t$ and $J_h$, and that $J_h$ is almost zero: $|J_d| \gg |J_t| > |J_h| \sim 0$.

4.2 Possible magnetic structures

Next, we discuss expected magnetic structures of the MLL antiferromagnet with $J_d$, $J_t$, and $J_h$. As mentioned in the introduction, the theoretically proposed ground state of the MLL Heisenberg antiferromagnet with $J_d = J_t = J_h$ is the coplanar magnetic structure shown in Fig. 2(a); here, the neighboring spins connected by $J_d$, $J_t$, and $J_h$ form 90°, 120°, and 150° angles, respectively. In every $J_t$ triangle, three spins form a 120° arrangement, as in the triangular lattice. The vector chirality on the $J_t$ triangle changes its sign alternately. Thus, we call this spin structure the SVC1 (staggered vector chirality 1) order.

We consider the case where $|J_d| \gg |J_t| > |J_h|$ for $MgMn_3O_7 \cdot 3H_2O$. In the first approximation, we assume that the spins on every $J_d$ dimer are antiparallel, which results in a rotation of 120° spins on the $J_t$ triangle while keeping the staggered arrangement of $\kappa$, as shown in Fig. 9(a). This magnetic structure is called SVC2 (staggered vector chirality 2). In the mean-field approximation, the transition temperatures of SVC1 and SVC2 are calculated to be $T_{SVC1} = 2C(-J_t - \sqrt{3}J_h)$ and $T_{SVC2} = 2C(-J_d - J_t + J_h)$. Thus, SVC2 is more stable than SVC1 in the case where $|J_h| < 0.366|J_d|$. If $|J_d|$ is large enough, then SVC2 would be selected as the ground state.

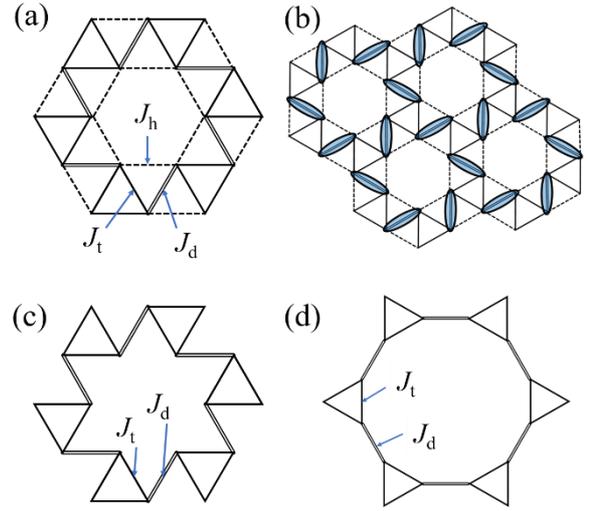

**Fig. 10** (a) MLL lattice with $J_d$, $J_t$, and $J_h$. (b) $J_d$ dimers on the MLL forming a kagomé lattice. (c) MLL in the limit where $J_h \to 0$. (d) Star lattice. The lattices in (c) and (d) are topologically equivalent.

There are two related magnetic structures that have the same energy as SVC2 but have different arrangements of $\kappa$: the PVC (positive vector chirality) order shown Fig. 9(b) with $\kappa = +1$ for every $J_t$ triangle and the NVC (negative vector chirality) order shown in Fig. 9(c) with $\kappa = -1$. Note that there is no correlation between the vector chiralities on neighboring triangles in a Heisenberg model with nearest-neighbor interactions. The presence of the three types of spin structure with different arrangements of $\kappa$ is similar to the case of the kagomé lattice antiferromagnet. When the $J_d$ dimer is regarded as a single spin, the MLL becomes a kagomé lattice, see Fig. 10(b). On the other hand, in the limit where $J_h = 0$, the MLL is equivalent to the star lattice, see Figs. 10(c) and (d). It is known that the vector chirality in the star lattice has a similar degree of freedom [18]. In these frustrated lattices, a secondary symmetry break may occur associated with the vector chirality degree of order.

At the moment it is not clear which kind of magnetic structure is realized, nor is the origin of the observed successive phase transitions in MgMn$_3$O$_7$·3H$_2$O. However, it is likely that one of the three above-mentioned magnetic structures is the ground state. The degeneracy in the vector chirality degree of freedom is usually lifted by additional magnetic exchange interactions and magnetic anisotropy. In the case of kagomé antiferromagnets, the Dzyaloshinskii–Moriya (DM) interaction stabilizes the PVC or NVC depending on its direction; the additional ferromagnetic (antiferromagnetic) next-nearest-neighbor interaction $J'$ stabilizes the SVC (PVC) [19, 20]. A similar situation must be realized in the MLL antiferromagnet. In MgMn$_3$O$_7$·3H$_2$O, DM interactions should exist on the $J_t$ and $J_h$ bonds (not on the $J_d$ bond) due to the absence of inversion symmetry. The observed weak-ferromagnetic behavior is probably a result of the canted-antiferromagnetic structure induced by the DM interactions. On the other hand, taking into account the next-nearest-neighbor interactions $J_h'$ in the $J_h$ hexagon, the energies of SVC and PVC (NVC) become different because the angles between the spins on the $J_h'$ bond between them are different. If these perturbations are weak, then a competition can induce successive phase transitions between these magnetic structures. To determine the magnetic structure, a neutron diffraction experiment is currently in progress.

On the other hand, there is an alternative scenario that explains the origin of the successive phase transitions. In the case of the Ising-spin triangular lattice antiferromagnet (TLA), a two-step phase transition is expected: the ordering of the $z$ component occurs at a high temperature, and that of the $xy$ component at a low temperature [21]. In the XY-spin TLA, an exotic chirality order without ordering of spin moments has been predicted [22]. In the MLL with a similar chiral degree of freedom, a phase transition for which spin chirality plays an important role may occur. In a frustrated spin system with a strong Ising anisotropy, successive phase transitions occur as a result of spin rearrangements [23], which is not applicable to the present system.

### 4.3 Possible magnetic plateau

Finally, we consider the origin of the multistep anomalies observed in the high-field magnetization. The two possibilities considered are magnetic-field induced phase transitions and magnetization plateaus. In the former case, there are four magnetic phases separated by the three critical fields of $\mu_0 H_{s1} = 33$ T, $\mu_0 H_{s2} = 40$ T, and $\mu_0 H_{s3} = 51$ T, where the $dM/dH$ curve at 1.3 K shows peaks. Successive phase transitions are expected as a result of the competition between almost-degenerate magnetic phases in the MLL antiferromagnet.

In the latter case, taking into account thermal effects, the dips in the $dM/dH$ curve at $\mu_0 H_{p1} = 35.7$ T and $\mu_0 H_{p2} = 45.1$ T correspond to the centers of two small magnetization plateaus. Moreover, a third, large plateau exists at $\mu_0 H_{p3} > 60$ T. The values of the magnetization at $H_{p1}$ and $H_{p2}$ are close to the fractional values of $M/M_s = 3/8$ and $1/2$, respectively, and that at $H_{p3}$ may approximate $M/M_s = 3/4$. This concurrence with simple fractional values suggests that a series of magnetization plateaus occur. It is well established that there is a simple relation between the fractional value of magnetization $M/M_s$ and the size of the magnetic unit cell n. According to the Oshikawa–Yamanaka–Affleck condition [24]: $6n \times S(1 - M/M_s)$ = integer, where $6n$ is the number magnetic ions in the magnetic unit cell and $n = 1$ for the crystallographic unit cell of the present compound. However, the observed series of plateaus are not realized for $n = 1$ and larger cells should be considered. Minimum cells contain 48 ($n = 8$), 12 ($n = 2$), or 24 ($n = 4$) Mn ions for $M/M_s = 3/8$, $1/2$, and $3/4$, respectively. If all of these plateaus come from a single unit cell, it must be assumed to be a large one of $n = 8$.

In the theoretical calculation for a $S = 1/2$ MLL antiferromagnet, it has been predicted that the $M/M_s = 1/3$ and $2/3$ plateaus emerge in the case of a large $J_d$ [18]. On the other hand, for a $S = 1/2$ star-lattice antiferromagnet ($J_h = 0$ in the MLL), plateaus with $M/M_s = 1/3$, $7/9$, and $8/9$ are theoretically predicted [25]. Our values are different from these. This difference may originate from two facts: our system is not quantum but close to classical with $S = 3/2$, and there must be additional terms other than the nearest-neighbor couplings in the Hamiltonian. To clarify the possibility of magnetization plateaus in MgMn$_3$O$_7$·3H$_2$O, further development of the theoretical investigation is needed.

### 5. Summary

We have synthesized the frustrated magnet MgMn$_3$O$_7$·3H$_2$O via a hydrothermal route and investigated its crystal structure, magnetism, and thermodynamic properties. We have shown that it is a good model compound for the maple-leaf-lattice antiferromagnet. In addition, MgMn$_3$O$_7$·3H$_2$O exhibits successive phase transitions at $T_{N1}$ = 5 K and $T_{N2}$ = 15 K. Under a high field, it displays multiple magnetic-field induced transitions, suggesting the presence of a series of magnetization plateaus. Judging from the relationship between the crystal structure and the magnetic interactions, a MLL lattice with $|J_d| \gg |J_t| > |J_h|$ is realized in which vector chirality plays a role in selecting the ground state from nearly degenerate spin orders. The MLL antiferromagnet will provide us with a unique platform to study frustrated magnetism.

## Acknowledgements

This work was partially supported by the Core-to-Core Program for Advanced Research Networks given by the Japan Society for the Promotion of Science (JSPS). Y.H. would like to acknowledge the support from the Motizuki Fund of the Yukawa Memorial Foundation.

----------------------------------------------------------------